\documentclass[12pt]{iopart}
\usepackage{epsfig}
\usepackage{amssymb}

\newcommand{\vek}[1]{{\mbox {\bfseries #1}}}
\newcommand{\vekt}[1]{{\mbox {\boldmath $#1$}}}
\begin{document}

\title[Q-LET --- Application to SN2003es]{Q-LET --- Quick Lensing
  Estimation Tool\\ 
An application to SN2003es}
\author{Christofer Gunnarsson}
\address{Stockholm University, AlbaNova University Center,
  Fysikum, SE-106 91 Stockholm, Sweden}
\ead{cg@physto.se}

\begin{abstract}
Q-LET is a FORTRAN 77 code that enables a quick estimate of the
gravitational lensing effects on a point- or an extended
source. The user provided input consists of the redshifts, angular positions
relative to the source, 
mass or velocity dispersion estimate and halo type for the
lens galaxies. The considered halo types are the Navarro-Frenk-White
and the
Singular Isothermal Sphere. The code uses the so-called multiple lens-plane
method to find the magnification and intrinsic shape of the
source. This method takes into account the multiple deflections
that may arise when several mass accumulations are
situated at different redshifts close to the line-of-sight.

The Q-LET code is applied to the recently disco\-vered super\-nova,     
SN2003es, which is likely to be of Type Ia as its
host galaxy is classified as an elliptical. We find that SN2003es is
likely to have been 
significantly magnified by gravitational lensing and that this should
be considered in high-$z$ studies if this SN is to be used to determine the
cosmological parameters.

Q-LET was motivated by the supernova searches, where lensing can be a
problem, but it can also be applied to any simple lens system
where a quick estimate is wanted, e.g.~the single lens case. 
\end{abstract}


\submitto{Journal of Cosmology and Astroparticle Physics, JCAP}

\maketitle
\section{Introduction}
The recent advance in modern cosmology due to new and
improved telescopes and techniques seems to overthrow the long-lasting
paradigm or hope of an Einstein-de Sitter (EdS) universe. From the
Cosmic Microwave 
Background (CMB) \cite{wmap,acbar,cbi}, along with studies of Large Scale
Structure (LSS), \cite{2df} and the Lyman $\alpha$ forest
\cite{lyalpha1,lyalpha2} (Ly$\alpha$)
etc.~it is inferred
that the universe consists of roughly 30 \% matter (Cold Dark Matter,
CDM and ordinary matter)
and 70 \% Dark Energy. Furthermore, measurements of the apparent magnitude
of Type Ia supernovae (SNe) have been used by two collaborations
\cite{scp,highz} to estimate the values of the total matter energy density 
$\Omega_{\rm M}$, which includes CDM, and the density in Dark Energy, 
e.g.~$\Omega_{\Lambda}$.  
Type Ia SNe are
believed to be so-called standard candles, i.e.~they show a very small
scatter in
intrinsic luminosity after empirical corrections
\cite{snscattcorr1,snscattcorr2}, and
by measuring the apparent magnitudes, the 
cosmology-dependent luminosity 
distance can be inferred. The
supernovae look 
fainter than they would in an EdS universe favouring both
a non-zero $\Omega_{\Lambda}$ and $\Omega_{\rm
M}\neq 1$. Combined with CMB data from Boomerang and MAXIMA
\cite{boomer,maxima}, they  
end up at approximately the same values as the above-mentioned CMB, LSS
and Ly$\alpha$ measurements.   
However, there is a number of potential systematic errors that can
contaminate the measurements. First, inter- and/or intra-galactic dust
may obscure the light from the SNe and make them look fainter. This
has been investigated in e.g.~\cite{edvariel1,aguirre1,aguirre2}. It is
also still unclear whether there is any evolution of the SN brightness with
redshift. Many potential sources of evolution have been proposed, such
as differing progenitor composition \cite{umeda,hoflich,domingez} or
host galaxy morphology (and thereby redshift) dependence 
\cite{snevol} just to name a few. Furthermore, if 
photons can oscillate into axions, the SNe will also look
fainter due to the reduced number of received photons. This has been
addressed e.g.~in reference \cite{edvard1}. A fourth
possible contaminant is magnification or de-magnification by
gravitational lensing. Having a large sample of SNe this can be
corrected for statistically \cite{edvard2}, but when studying
single SNe, 
the lensing effects need a careful treatment. This was the case
with the farthest known SN so far, SN1997ff, at a redshift of 1.7.
The vicinity of the line-of-sight to this SN was unusually dense in
galaxies and 
modelling of the individual lens galaxies was required. This was done
by \cite{lewiba,cgedvariel,riess1,benitez} who conclude a large
possible magnification depending on the lens masses and
concentrations. Ben\'{\i}tez et al.~\cite{benitez} conclude a 
magnification of 0.34$\pm$0.12 mag.  

In this paper we present the well-known multiple lens-plane method
applied to the calculation
of the magnification 
of any point- or extended source. Section \ref{sec:lensing} describes
this method 
in a general 
way. Furthermore, we model the lenses as spherically
symmetric 
Navarro-Frenk-White (NFW) or Singular Isothermal Sphere (SIS) halos and
the lensing properties of these are described in 
sections \ref{subs:nfw} and \ref{subs:sis}. For more information on under
which circumstances each halo type is appropriate, see
e.g.~reference \cite{liost}. The  
method and models are manifested in a 
\texttt{FORTRAN 77} 
program, Q-LET, obtainable from the author upon request or it can be
downloaded at \texttt{http://www.physto.se/\~{}cg/qlet/qlet.htm}.
The program is 
described in the text in section \ref{sec:program}. 
In reference \cite{cgedvariel}, an early version of Q-LET
was used to estimate the magnification of the abovementioned SN1997ff.
Section \ref{sec:ramone} describes the application of the code to the
recently discovered supernova SN2003es. 
     
\section{Multiple light deflection}
\label{sec:lensing}

This section introduces the multiple lens-plane method and gives the
important equations needed in the study of multiple light
deflection. For more details, see reference \cite{schneider}. We also
describe the lensing properties of the NFW and SIS halo profiles. 

\subsection{The multiple lens-plane method}
\label{subs:mult}
\begin{figure}
\center{\epsfig{file=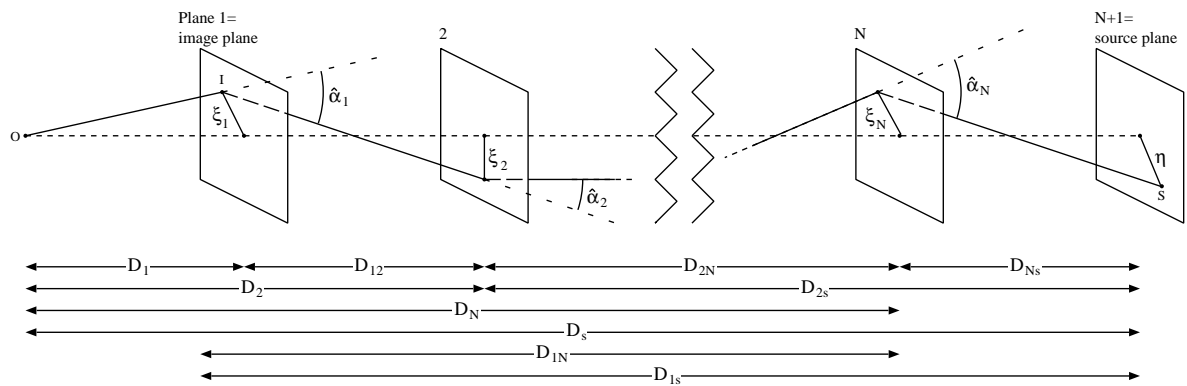,width=1\textwidth}}
\caption{Schematic picture of a multiple lens-plane situation with
observer $O$, image $I$ and source $S$.}
\label{fig:mullens}
\end{figure}  

We start by projecting the mass of each massive object (e.g.~galaxy)
onto a plane at the 
respective object redshift. In the $i$:th plane, this gives a surface mass
density $\Sigma_i(\vekt{\xi}_i)$, where $\vekt{\xi}_i$ is the impact
parameter of the light-ray in each plane (see
figure \ref{fig:mullens}). The justification 
for this projection is that the light is deflected only in the
very vicinity of the lens in most cases and then travels practically
unimpeded until it reaches the observer or another lens, situated far
away from the preceding one.   

\subsubsection{The lens equation}
From figure \ref{fig:mullens}, where most quantities are defined, we see
that the position of the light-ray 
in each plane can be obtained recursively from the observed position
$\vekt{\xi}_1$ through 
\begin{equation}
\label{eq:leneq}
\vekt{\xi}_j=\frac{D_j}{D_1}\vekt{\xi}_1-\sum_{i=1}^{j-1}D_{ij}\vekt{\hat{\alpha}}_i(\vekt{\xi}_i),
\end{equation}
where $D_{ij}=D(z_i,z_j)$ is the angular diameter distance between
redshifts $z_i$ and $z_j$, $D_i=D(0,z_i)$ and $D_{is}=D(z_i,z_{\rm
source})$. With $N$ lens-planes we find the source position as
$\vekt{\eta}=\vekt{\xi}_{N+1}$. For convenience we use, in each plane, 
dimensionless quantities defined by
\begin{equation}
\label{eq:dimless1}
\vek{x}_i=\frac{\vekt{\xi}_i}{\xi_{0i}},
\end{equation}
where $\xi_{0i}$ is an arbitrary scale length in the $i$:th plane
which, if chosen as $D_i$,  
makes $\vek{x}_i$ the angular impact parameter in each
plane. Furthermore, the \emph{convergence} is 
\begin{equation}
\label{eq:conv}
\kappa_i(\vek{x}_i)\equiv\frac{\Sigma (\xi_{0i} \vek{x}_i)}{\Sigma_{\rm crit}},
\end{equation}
where (in geometrised units)
\begin{equation}
\Sigma_{\rm crit}=\frac{D_s}{4\pi D_i D_{is}}
\end{equation}
is called the \emph{critical density} due to its relation to the
ability of a lens to produce multiple images. 
If we also define $\vekt{\alpha}_i=\vekt{\hat{\alpha}}_i D_{is}/D_s$, 
the \emph{scaled deflection
potential} $\psi_i(\vek{x}_i)$, needed for the magnification and
defined through $\vekt{\alpha}_i=\nabla
\psi_i$ becomes
\begin{equation}
\label{eq:scdefpot}
\psi_i(\vek{x}_i)=\frac{1}{\pi}\int_{\mathbb{R}^2}
\kappa_i(\vek{x}')\ln |\vek{x}_i-\vek{x}'|\rmd^2x'.
\end{equation}
In these units where $\vek{y}\equiv \vek{x}_{N+1}$ is the source position, the lens
equation is given by
\begin{equation}
\label{eq:leneq2}
\vek{y}=\vek{x}_1-\sum_{i=1}^N \vekt{\alpha}_i(\vek{x}_i).
\end{equation} 

\subsubsection{The magnification}
\label{subs:magn}
We now want to find the magnification as a function of image position.
Let $A(\vek{x}_1)$ denote the Jacobian matrix of the lens equation
\begin{equation}
A(\vek{x}_1)\equiv \frac{\partial \vek{y}}{\partial \vek{x}_1}.
\end{equation}
Furthermore we need 
\begin{displaymath}
U_i\equiv \frac{\partial \vekt{\alpha}_i}{\partial
\vek{x}_i}=\left(\begin{array}{cc}
  \frac{\partial \alpha_{i1}}{\partial x_{i1}} & \frac{\partial \alpha_{i1}}
  {\partial x_{i2}} \\
 {} & {} \\
  \frac{\partial \alpha_{i2}}{\partial x_{i1}} &  \frac{\partial \alpha_{i2}}
  {\partial x_{i2}} \end{array}\right)=\left(\begin{array}{cc}
  \psi_{i11} & \psi_{i12} \\
  \psi_{i21} &  \psi_{i22} \end{array}\right)=
\end{displaymath}
\begin{equation}
=\left(\begin{array}{cc}
  \kappa_i +\gamma_{i1} & \gamma_{i2} \\
  \gamma_{i2} &  \kappa_i -\gamma_{i1} \end{array}\right),
\label{eq:ui}
\end{equation} 
where 
\begin{equation}
\psi_{ikl}=\frac{\partial^2 \psi_i}{\partial x_k \partial x_l},
\label{eq:psidef}
\end{equation} 
$\nabla^2_{\vek{x}_i}\psi_i=2\kappa_i(\vek{x}_i)$,
$\gamma_{i1}=(\psi_{i11}-\psi_{i22})/2$ and
$\gamma_{i2}=\psi_{i12}=\psi_{i21}$. 
This gives $A$ as
\begin{equation}
A
={\mathbb I}-\sum_{i=1}^N
\frac{\partial \vekt{\alpha}_i}{\partial \vek{x}_1}={\mathbb I}-\sum_{i=1}^N
\frac{\partial \vekt{\alpha}_i}{\partial \vek{x}_i}\frac{\partial
\vek{x}_i}{\partial \vek{x}_1}={\mathbb I}-\sum_{i=1}^N U_i A_i,
\label{eq:amult}
\end{equation}
where ${\mathbb I}$ is the 2$\times$2 unit matrix. The $A_i$:s can be
found by recursion and by noting that $A_1={\mathbb I}$;
\begin{equation}
A_j={\mathbb I}-\sum_{i=1}^{j-1} \beta_{ij} U_i A_i.
\label{eq:a}
\end{equation}
Finally, the magnification is given by
\begin{equation}
\mu=\frac{1}{{\rm det}A}=(A_{11}A_{22}-A_{12}A_{21})^{-1}.
\label{eq:mag}
\end{equation} 
Negative values of $\mu$ indicates images of reversed parity relative
to the unlensed image.

\subsection{Properties of the NFW halo profile}
\label{subs:nfw}
The use of the Navarro-Frenk-White model calls for a brief description
of its lensing properties. 

Before projection onto the lens plane, the NFW profile is given by
\begin{equation}
\rho_{\rm NFW}(r)=\frac{\rho_s}{\frac{r}{r_s}\left(
1+\frac{r}{r_s}\right)^2},
\label{eq:nfw}
\end{equation}
where $\rho_s$ and $r_s$ are scale parameters that can be determined once 
e.g.~the mass is specified, $r$ is the radial co-ordinate.    
The projected surface mass density will be circularly symmetric and we
no longer need vector notation on the impact parameter but instead
choose $\xi$ as the radial co-ordinate with the origin at the lens
centre. If we choose $\xi_0=r_s$ (cf.~equation (\ref{eq:dimless1})) and
define $\kappa_s=\rho_s r_s/\Sigma_{\rm crit}$, then 
\begin{equation}
\kappa(x)=\frac{2\kappa_s}{x^2-1}f(x)
\label{eq:nfwkappa}
\end{equation}
where
\begin{equation}
f(x)=\left\{ \begin{array}{cc}
 1-\frac{2}{\sqrt{x^2-1}}\arctan \sqrt{\frac{x-1}{x+1}} & ,x>1 \\
1-\frac{2}{\sqrt{1-x^2}}{\rm arctanh} \sqrt{\frac{1-x}{1+x}} & ,x<1 
\end{array} \right. ,
\label{eq:fofx}
\end{equation}
and $\kappa(1)=2\kappa_{\rm s}/3$.
The deflection angle is
\begin{equation}
\alpha(x)=4\kappa_s\frac{g(x)}{x}
\label{eq:alfanfw}
\end{equation}
where
\begin{equation}
g(x)=\ln \frac{x}{2}+ \left\{ \begin{array}{cc}
 \frac{2}{\sqrt{x^2-1}}\arctan \sqrt{\frac{x-1}{x+1}} & ,x>1 \\
\frac{2}{\sqrt{1-x^2}}{\rm arctanh} \sqrt{\frac{1-x}{1+x}} & ,x<1 \\
1 & ,x=1 \end{array} \right. 
\label{eq:gofx}
\end{equation}       
and the magnification will be given by 
\begin{equation}
\mu(x)=\left[\left(1-\frac{\alpha(x)}{x}\right)\left(1+\frac{\alpha(x)}{x}-
  2\kappa(x)\right)\right]^{-1}. 
\label{eq:mag5}
\end{equation}
A flaw of this model is that the total mass diverges when integrated
out to infinity. However, this is not too serious since the deflection
angle only is sensitive to the mass inside the impact radius, and the
magnification is sensitive to this mass and the convergence at this
point (for circularly symmetric lenses). Thus all mass outside the
impact radius is unimportant. 

A single NFW halo gives either one or three images, where the primary
image has $\mu\geq 1$, the secondary $\mu<0$ and the tertiary $\mu>0$. 

\subsection{Properties of the SIS halo}
\label{subs:sis}
The Singular Isothermal Sphere model, also used here, is based on
the assumption that the dark matter in the halo behaves as particles
in an ideal gas trapped in their gravitational potential. The gas is
assumed to be in thermal equilibrium and the resulting
density profile before projection (in geometrised units) is
\begin{equation}
\rho_{\rm SIS}(r)=\frac{\sigma^2_v}{2\pi r^2},
\label{eq:sis}
\end{equation}
where $\sigma_v$ is the line-of-sight velocity dispersion of the particles.
If we choose 
\begin{equation}
\xi_0=4\pi \sigma_v\frac{D_d D_{ds}}{D_s}
\label{eq:xi0sis}
\end{equation}   
the equations for convergence, deflection angle, intrinsic source
position and magnification 
become very simple for this model: 
\begin{equation}
\kappa(x)=\frac{1}{2x},
\label{eq:kappasis}
\end{equation} 
\begin{equation}
\alpha(x)=\frac{x}{|x|}
\label{eq:alfasis}
\end{equation}       
and
\begin{equation}
y(x)=x-\frac{x}{|x|}
\label{eq:ysis}
\end{equation} 
and
\begin{equation}
\mu(x)=\frac{|x|}{|x|-1}.
\label{eq:musis}
\end{equation} 
As with the NFW model, the SIS will give an infinite mass when
integrated to infinity but the argument above regarding deflection
angle and magnification also holds here.
When considering a single SIS halo it
can give either one or two images depending on the
impact parameter. Primary images have $\mu\geq 1$ and secondary have $\mu<0$.

\section{The Q-LET code}
\label{sec:program}
The Quick Lensing Estimation Tool, Q-LET is a
\texttt{FORTRAN 77} code written in order to quickly be able to
estimate the 
lensing effects on a point source or an extended object. 
The structure of the code is as follows
\begin{itemize}
  \item Read the input datafile and sort the planes 
  \item Check if parameter values are allowed
  \item Present a choice between a point- or an extended
  source with elliptical image shape, alternatively a square grid (see
  section \ref{subs:output}) 
  \item Compute magnifications and positions of light rays
  \item Present a choice of how to output the results
\end{itemize}
A used supplied datafile contains the cosmological
parameters, the source redshift and position and the lenses' redshifts,
positions, velocity dispersions \emph{or} masses and finally their halo
type. The mass should be ${\rm M}_{200}$, i.e.~the mass within the
radius ($\equiv r_{200}$) within which the average energy density is
200 times the critical 
density at the lens' redshift. 

A point source must be put at the origin. For the case of an elliptical image,
the origin will be put at the source centre and the ellipse will be
centered at user specified co-ordinates which should be taken as (0,0)
if an elliptical image only is studied. The possibility of putting the
ellipse off-origin is useful to study
e.g.~a SN that is offset from its host galaxy centre and the galaxy
shape is to be investigated. The SN is put at the origin and the
host galaxy central co-ordinates are given as input. If a
square grid (see section \ref{subs:output}) is used, it will be centered at
$(0,0)$. The 
angular positions of the lensing galaxies should of course be assigned
with respect to
this origin and should be given in arcseconds. The velocity
dispersions and masses should be given in ${\rm km}\;{\rm s}^{-1}$ and
${\rm M}_{\odot}$ respectively. See section \ref{subs:input} for an
example of an input file. 

If only ${\rm M}_{200}$ is given for a SIS lens, this mass will be
used to calculate the corresponding velocity dispersion since
$\sigma_{\rm v}$ is the parameter used to determine the lensing
strength for  
this halo type. If, on the other hand, only $\sigma_{\rm v}$ is given
for a NFW lens, this velocity dispersion will be used to calculate the
corresponding ${\rm M}_{200}$ for a {\bf SIS} lens and assuming ${\rm
  M}_{200}$ is the same for the NFW halo\footnote{See
  section \ref{subs:ramdisc} for a discussion of the validity of this.}. This
is done since the 
lensing properties of the NFW halo depends upon ${\rm M}_{200}$
in our calculations. If, for some reason, both the mass and velocity
dispersion are given, the mass (velocity dispersion) will be ignored
for the SIS (NFW) lens. 

All distances are computed using the filled beam approximation and it
is also with respect to an homogeneous universe that the magnification
is given.

\subsection{Parameter restrictions}
\label{subs:restr}
Due to computational complications (and perhaps lack of connection to
reality) there are some approximate restrictions on the
input parameters. However, these restrictions do not severely limit
the usefulness of the code as can be seen in table
\ref{tab:restr} where the allowed parameter ranges are
presented. Furthermore, some of them are not very strict and depends
on the values of the other parameters making these ranges
approximate only. Especially extremely low or high Hubble parameters ($\sim
0.1$ or $\sim 1.5$ might lead to problems. 
\begin{table}
\begin{center}
 \begin{tabular}{lc} 
 \br
 Source redshift & $0<z_{\rm s}\lesssim 20$ \\
 Lens redshift & $0<z_{\rm l}<{\rm min}[z_{\rm s},5]$ \\
 Lens mass $[{\rm M}_{\odot}]$ & $10^6\lesssim {\rm M}_{200}\lesssim
 10^{17}$ \\ 
 Lens velocity disp. $[{\rm km}\;{\rm s}^{-1}]$ &  5$\lesssim
 \sigma_{\rm v} \lesssim 5000$ \\\mr
 Hubble parameter &
 $0<h<2$ \\
 Mass energy density & $0\leq \Omega_{\rm
 m}\leq 10$ \\ 
 Energy density in cosmological const. & $-10\leq
 \Omega_{\Lambda} \leq 10$ \\\mr
 Combined restrictions & $\Omega_{\rm m}+\Omega_{\Lambda}=1$ or \\
 when using NFW halo & $\Omega_{\rm m}<1$ and $\Omega_{\Lambda}=0$ \\
  \br
 \end{tabular}
\caption{Approximate parameter restrictions in the code. The Hubble
  parameter and 
  energy densities are
  given in units of 100 ${\rm km}\;{\rm s}^{-1}\;{\rm Mpc}^{-1}$ and
  critical energy density, $\rho_{\rm crit}$ respectively.}
\label{tab:restr}
\end{center}
\end{table} 

\subsection{Input}
\label{subs:input}
The information needed by Q-LET to be able to compute the
magnification and deflection of a light ray is the following: the values of the
cosmological parameters, i.e.~$h$, $\Omega_{\rm m}$ and
$\Omega_{\Lambda}$, the source redshift and position, the lenses'
redshifts and positions, their velocity dispersion or mass and their
halo types (NFW or SIS). This information is collected in an input
datafile which could have the following appearance:
\begin{verbatim}    
   0.7     0.3     0.7
   1.7       0       0       0         0  xxx	
 0.923    -4.1    3.01     138         0  nfw
 0.655     2.6   -1.11       0      1e12  sis
 0.322   -3.32    5.11       0   9.87e10  nfw
 1.453   -2.32    -0.4   180.0         0  sis
 0.011   -2.34    2.32       0     6.5e9  nfw
 0.234    -4.1    3.01     190         0  sis
 0.254   -5.23   -2.43   100.3         0  nfw
\end{verbatim}
The entries are:\\
\begin{tabular}{rrrrrr}
$h_0$ & $\Omega_{\rm m}$ & $\Omega_{\Lambda}$ & - & - & - \\
$z_{\rm source}$ & $x_{\rm source}$ [$''$] & $y_{\rm source}$ [$''$] &
  0 & 0 & 
  \verb2xxx2 \\
$z_{{\rm lens}_1}$ & $x_{{\rm lens}_1}$ [$''$] & $y_{{\rm lens}_1}$ [$''$] &
  $\sigma_{{\rm v}_1}$ [${\rm km}\;{\rm s}^{-1}$] & ${\rm M}_{{200}_1}$ [${\rm
      M}_{\odot}$] & halo type$_1$ (NFW/SIS)\\
$\vdots$ & $\vdots$ & $\vdots$ & $\vdots$ & $\vdots$ & $\vdots$ \\
$z_{{\rm lens}_n}$ & $x_{{\rm lens}_n}$ [$''$] & $y_{{\rm lens}_n}$ [$''$] &
  $\sigma_{{\rm v}_n}$ [${\rm km}\;{\rm s}^{-1}$] & ${\rm M}_{{200}_n}$ [${\rm
      M}_{\odot}$] & halo type$_n$ (NFW/SIS)\\
\end{tabular}\\
{}\\
For
easier handling of the data you should put '0' in the fourth
and fifth columns and e.g.~\verb2---2 in the sixth column of row
2. Remember to put the source at (0,0) and of course at 
the highest redshift. 
The following rows (up to 199 more rows) should
contain data on the lensing halos according to the example above.
 
\begin{figure}
\center{\epsfig{file=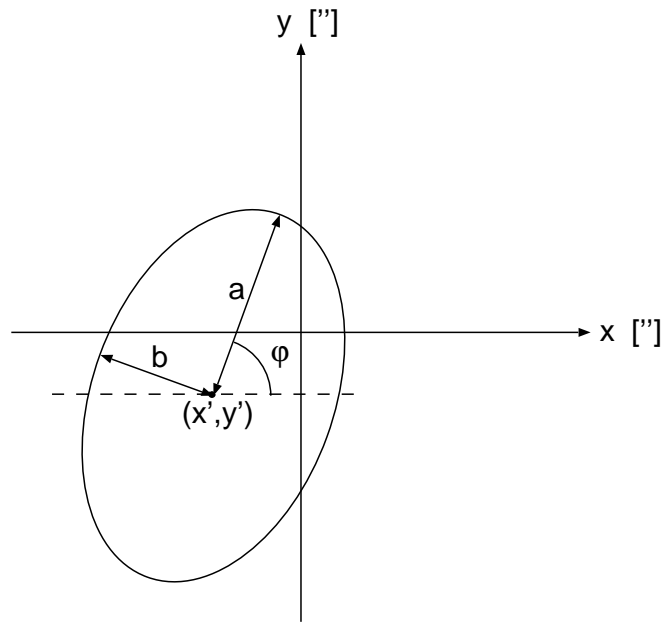,width=0.55\textwidth}}
\caption{Definitions of major, $a$, minor, $b$, axes, rotation
  angle, $\varphi$ and offset $x'$ and $y'$. Units: arcseconds (axes
  and offset) and degrees (rotation angle).} 
\label{fig:ell}
\end{figure}

The program starts by asking for this datafile. The default name is 
'\verb2qdata.qlet2'. In the next step, a question is posed on whether
you want to study a point source, an elliptical image or a 20$''\times 20
''$ square grid\footnote{Can be used to trace an arbitrary image shape back to
  the source-plane to see the intrinsic shape.}. 
If you choose an elliptical image, you will be asked for another input
file with major and minor axes, rotation angle and possible offset
from the source (see
figure \ref{fig:ell}). The default name of this input file 
is '\verb2elldata.ell2' and consists of one line only. An example
input file is:\\
\verb+0.8 0.3 72.0 -0.47 0.12+\\
The inputs are $a\ \ b$ in arcseconds, $\varphi$ in degrees and the
offset $x'$ and $y'$ in arcseconds.

\subsection{Output}
\label{subs:output}
If a point source is
chosen, it is placed at (0,0) and traced back to the source plane and
the magnification is computed. This is printed on the screen, and it
is also possible to obtain the output ($x$-, $y$-position and
magnification) in a small file. 
If a magnification $\mu<1$ is obtained, it is not a primary image that
has been found, i.e.~muliple imaging has occurred. 
In the case of an elliptical image, 
the user is asked to provide the number of the plane wished to be output
in a file. You can choose any plane between 2 and the number of the
source plane if you would like to study the intermediate changes of
the ray bundle. Each row in the output file refers to a specific ray
and could look like:\\
\verb+8395.  0.2131891  0.8527565 -2.296262  0.633064  2.873963+\\
The entries are:\\
\begin{tabular}{rrrrrr}
No.~of the ray & $x_1$ [$''$] & $y_1$ [$''$] & $x_i$ [$''$] & $y_i$
[$''$] & magnification,\\
\end{tabular}\\
where subscript 1 refers to the image plane and $i$ refers to the
chosen plane. The file will
contain a quite large amount of rays in order to see the shape better
(of order 1-5$\times 10^4$, depending on the size of the ellipse).    
To see the result, just plot the file with the desired columns.

When a square grid is chosen, the output file will look exactly like
in the ellipse-case, the difference is that the grid is square and
pre-defined to be 20$''\times20''$. This grid can be used if the image(s)
has (have) a shape not described by an ellipse. When trying to
estimate the intrinsic shape of an object like this, you plot the 4:th
and 5:th columns making cuts using the 2:nd and 3:rd columns as to only
include the rays making up the image(s). 
If instead wishing to see e.g.~what an intrinsically
elliptical source looks like in the image plane, you make the
cut using the 4:th and 5:th columns, even though the grid in the
source plane is not evenly distributed it will be possible as long as
it covers the area where the cut is made. 

\section{Example: application to SN2003es}
\label{sec:ramone}
SN2003es was found at a redshift of 0.968 in the
Hubble Deep Field North (HDF-N) in
April 2003 by the GOODS/HST Transient Search, as part of the HST
Treasury program \cite{goodshst-t}. As the host galaxy seems
to be an elliptical, it is most
likely a Type Ia SN \cite{Iaell}. Since these are used as standard
candles to determine
the redshift-distance relation in the quest for the cosmological
parameters, it is very important to study the potential magnification
by gravitational lensing. There 
are several possible lens 
candidates close to the line-of-sight of SN2003es so a close inspection
is necessary.
\subsection{Host and lensing galaxy data}
\label{subs:ramonedata}
The data on the host and lens galaxies has been taken from three different
references; \cite{fernan,gwyn,cohen}. Since the
redshift information 
differ significantly in the different catalogs, we used the principle to
take the spectroscopic redshifts when available and the 
photometric redshifts otherwise. If two catalogs had only photometric
redshifts, the most recent result was used. However, as a worst
case scenario regarding the amount of magnification, we also
computed the magnification including four galaxies from one of the older
catalogues which, in the most recent catalogue, were determined to be
background galaxies at $z>z_{\rm source}$. Hereafter we call these GH after
Gwyn and Hartwick of reference \cite{gwyn}.

To estimate the velocity dispersion that is needed in the lensing
calculation, we took the observed luminosities and best fit galaxy
morphology from \cite{fernan} and performed cross-filter K-corrections
to rest frame 
$B_J$ using galaxy templates provided by the SNOC package
\cite{snoc}. The Tully-Fisher (T-F) relation for spirals 
(S{\texttt xx}) and
Faber-Jackson (F-J) relation for ellipticals (Ell) and irregulars
(Irr) were used to derive an
approximate velocity dispersion. 
The relations are given by
\begin{equation}
\frac{\sigma_{\rm v}}{\sigma_*}= 10^{0.088(M_*-M)}\ \ \ \ \ ({\rm T-F})\\
\label{eq:tf}
\end{equation}
and
\begin{equation}
\frac{\sigma_{\rm v}}{\sigma_*}= 10^{0.1(M_*-M)}\ \ \ \ \ \ \ \ ({\rm F-J}),
\label{eq:fj}
\end{equation}
where $\sigma_*$ is the normalisation of the velocity dispersion, $M$
is the absolute magnitude in $B_J$ and $M_*$ is a typical magnitude
taken to be $M_*=-19.53+5\log h$ \cite{peebles}.

Within a radius of 10$''$ we found 9 lensing galaxies excluding GH and
13 lenses including it. The redshifts, velocity dispersions, positional
data and galaxy types can be found in table \ref{tab:ramgalinf}.
\begin{table}
\begin{center}
 \begin{tabular}{clrrrc} 
 \br
 Cat.~no\footnotemark
 & \multicolumn{1}{c}{$z$} & $x$-pos.~$['']$ & $y$-pos.~$['']$ &
 $\sigma_{\rm v}/\sigma_*$ 
 & Gal.~type\\
\mr
 526 & 0.120 & -3.74 & -8.60 & 0.136 & Scd \\
 539 & 0.120 & -2.55 & -7.93 & 0.112 & Scd \\
 549 & 0.952$^a$ & 1.62 & -8.29 & 0.924 & Ell \\
 622 & 0.440 & 4.76 & -2.59 & 0.332 & Scd \\
 623 & 0.960 & -6.80 & 2.58 & 0.334 & Scd \\
 631 & 0.321$^a$ & 1.69 & -0.52 & 0.460 & Irr \\
 662 & 0.511$^a$ & 2.71 & 2.88 & 0.557 & Irr \\
 693 & 0.600 & 2.52 & 6.85 & 0.402 & Ell \\
 700 & 0.040 & -1.52 & 9.28 & 0.098 & Sbc \\
\mr
\multicolumn{6}{c}{GH galaxies}\\
\mr
 582 & 0.904 & -9.73 & -0.62 & 0.401 & Scd \\
 607 & 0.776 & 1.64 & -3.07 & 0.431 & Sbc \\
 610 & 0.296 & 0.87 & -2.43 & 0.218 & Scd \\
 696 & 0.904 & 8.52 & 4.51 & 0.682 & Irr \\
\mr
\multicolumn{6}{c}{Host galaxy}\\
\mr
 619 & 0.968$^a$ & -0.475 & -0.710 & (1.290) & Ell \\
\br
\end{tabular}
\caption{Lensing galaxy data. The galactic centre position is given relative to the
  co-ordinates of SN2003es. The galaxy type is the best fit from
  reference \cite{fernan}. The galaxies with spectroscopically determined redshifts
  are superscripted $a$.}
\label{tab:ramgalinf}
\end{center}
\end{table} 
\footnotetext{Catalogue number of reference \cite{fernan}.}
\subsection{Results}
\label{subs:ramres}
\begin{figure}
\center{\epsfig{file=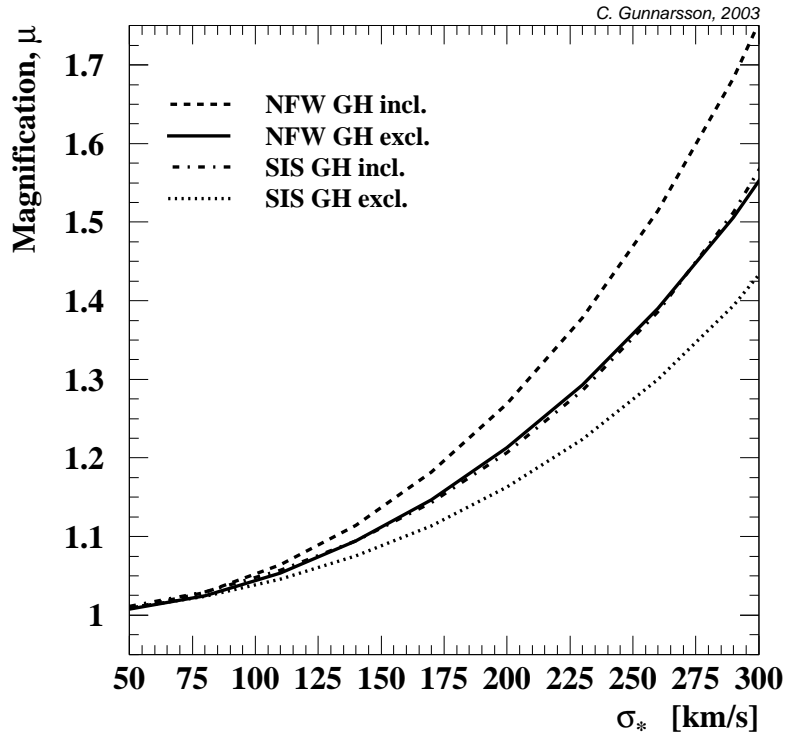,width=0.8\textwidth}}
\caption{Magnification of SN2003es as a function of the velocity
  dispersion normalisation.}
\label{fig:rammag}
\end{figure}

We used the Q-LET code to compute the magnification of SN2003es and also
to find the intrinsic shape of the host galaxy. This is done in order
to see that the observed elliptical shape is not contrived from an
intrinsically very complicated shape that is never observed in
reality. 
The normalisation of the velocity dispersion is quite poorly known
\cite{wilson} and 
we choose to present the magnification as a function of this
normalisation. For the host galaxy shape calculation we used a value
of $\sigma_*=170$ km/s, obtained as an average in
reference \cite{fischer} for a mixture of galaxy morphologies. However, the
shape is not distorted  
very much even with $\sigma_*$ as high as
300 km/s, although it is translated and more strongly magnified. 
The result on the magnification is found in figure \ref{fig:rammag}. We
have used models where all halos are assumed to be either of NFW or
SIS type. The figure also shows the result both when including the
four GH galaxies and when excluding them. As is seen, in the region
of $\sigma_*$
between 150 and 200 km/s, i.e.~around 170 km/s, the magnification
varies between roughly 1.08 and 1.27 depending on the the model and
the presence of the GH galaxies. This corresponds to a magnitude
shift of $0.1\lesssim \Delta m\lesssim 0.25$ mag, a significant
change that 
could dominate the uncertainty (the intrinsic magnitude spread in Type
Ia SNe is 
roughly between 0.1 and 0.2 mag \cite{hamuy}). Note also that higher
$\sigma_*$ have been  
obtained e.g.~in reference \cite{fukugita}, where a value for the SIS of
roughly 275 km/s was obtained for the dark matter in ellipticals. 

\begin{figure}
\center{\epsfig{file=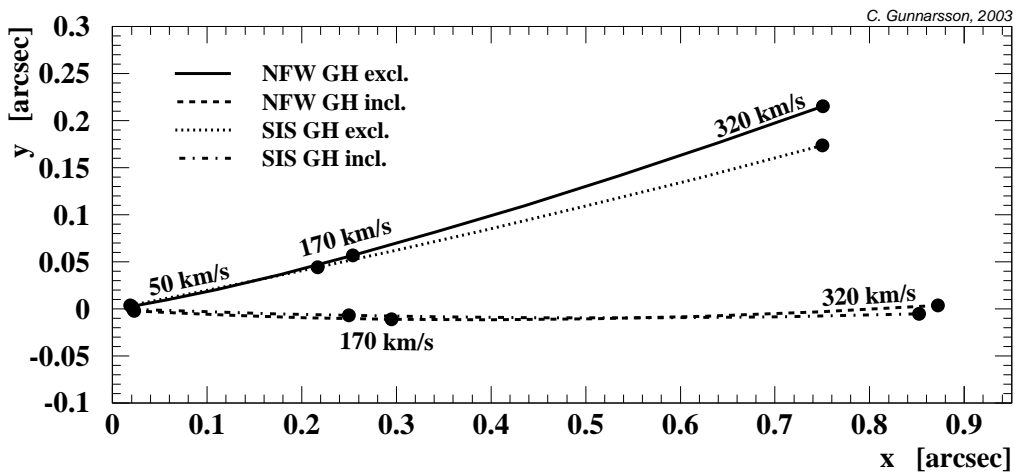,width=1\textwidth}}
\caption{The intrinsic (unlensed) position of SN2003es relative to the observed
  position at (0,0). The position will move along the curves from left to right
  with increasing $\sigma_*$ as indicated.}
\label{fig:move}
\end{figure}

In figure \ref{fig:move}, we display the intrinsic (unlensed) position
relative to 
the observed position at (0,0), depending on $\sigma_*$ and the model
(NFW/SIS and GH incl./GH excl.). We see that the shift is only weakly depending
on the halo type for this lens system under our assumptions.  

\begin{figure}
\center{\epsfig{file=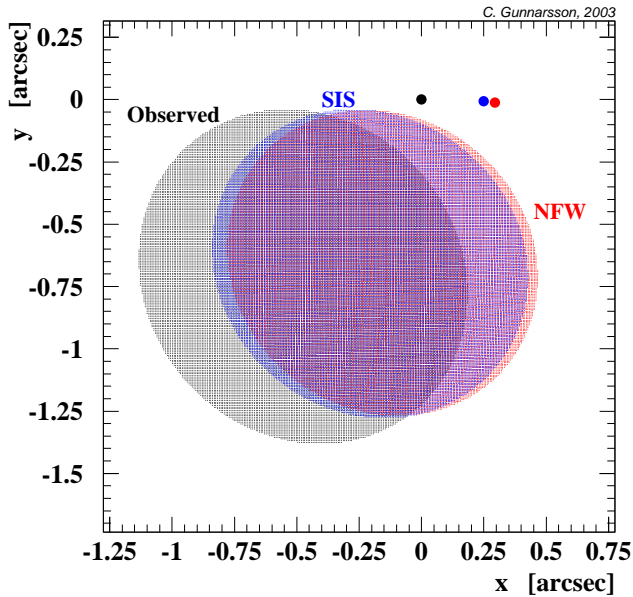,width=0.65\textwidth}}
\caption{The observed and intrinsic shapes of the host galaxy. The larger
  dots indicate the position of the supernova. The figure shows the small 
  difference in deflection between NFW and SIS halo models under our
  assumptions.}
\label{fig:ramell}
\end{figure}

Finally, we show the intrinsic shape of the host galaxy for a
$\sigma_*$ of 170 km/s \cite{fischer} in figure \ref{fig:ramell}. This plot
\emph{also} shows the small difference in deflection between NFW and SIS
halo models under our assumptions. The shapes were computed including the GH
galaxies. The axis ratios of both the NFW and SIS case are roughly 0.9,
making them perfectly normal E1 galaxies so no constraints on the
magnification or model can be inferred from the host galaxy shape.

\section{Discussion --- Q-LET \& SN2003es}
\label{subs:ramdisc}
One of the approximations that may be criticised is the fact that we
extend the halos out to infinity. However, the lensing effect at large
impact radii is very 
small. For SN2003es, the lenses with large impact radii contribute
very little to the total magnification which is dominated by galaxies
no.~631 and 662.

Putting ${\rm M}_{200}$ of the NFW equal to that of the SIS for a
given velocity dispersion is not obviously a good approximation. The
most likely measured quantities are the luminosity or velocity
dispersion of the lens and the luminosity may be used to estimate a
velocity dispersion either via the Tully-Fisher or Faber-Jackson relation.
When a velocity dispersion is obtained, the best way would be to use
it directly on the NFW profile. However, as opposed to the SIS, the NFW
velocity dispersion depends on the radius. By following {\L}okas and
Mamon in reference \cite{lokas} it is possible to obtain the line-of-sight
velocity dispersion, $\sigma_{\rm los}$, for well resolved sources, or,
for small or distant 
sources, the ``aperture velocity dispersion'', $\sigma_{\rm ap}$,
which is the average over 
an aperture centered on the object. We calculated both these
quantities as a function of radius and found that $\sigma_{\rm ap}$
did not fall off dramatically. As an example, in a case where we put the SIS velocity
dispersion ($\sigma_{\rm v}^{\rm SIS}$) to 170 km/s to get the parameters for the NFW and then
compute the corresponding NFW velocity dispersions, $\sigma_{\rm ap}$
varies roughly by $+5$\% and 
$-15$\% around 170 km/s between radii $0.1r_{200}$ and $2r_{200}$. The
line-of-sight velocity dispersion varies more, $+10$\% to $-40$\% in
the same range. However, at a radius of \emph{roughly} $0.3r_{200}$,
there is a quite small scatter around the SIS value of less
than $\sim$15\%. This seems to be valid for a large range of redshifts
and velocity dispersions. It also seems valid in low $H_0$- and low
density universes. Thus in almost all cases it is possible to find a
radius at which $\sigma_{\rm los}\sim\sigma_{\rm ap}\sim\sigma_{\rm
  v}^{\rm SIS}$, with this radius being roughly equal to $0.3r_{200}$. 

Inverting the velocity dispersion-mass relation for the NFW is
quite time-consuming, and this, along with the ambiguity due to the
dependence of 
$\sigma_{\rm v}$ on radius and the fact that the difference seems not to
be significant, motivates using the simpler conversion via
the SIS. However, if this conversion is unsatisfactory, it is always
possible for the user to supply ${\rm M}_{200}$ instead, in which case
of course no conversion is needed.

For the irregular galaxies we have used the
Faber-Jackson relation which is valid for ellipticals. An
approximation which is difficult to say to what extent it affects
the results, and in which direction. Using the Tully-Fisher relation
on the irregulars would increase the magnification even further, but
at most by $\sim$15-20\%.

Whether the halo models we use are good descriptions of real halos is
still under debate. We are using spherically symmetric models, which
might be an oversimplification but the ellipticity of the dark matter
in the observed halos is in practice very seldom known (see
e.g.~reference \cite{elllens} for the effects of ellipticity in lensing).  

So, unless some new model is very different from
the SIS and NFW, the results for SN2003es is likely to remain.

\subsection{Conclusions}
\label{subs:ramconcl}
We have found that SN2003es is likely to have been significantly
magnified by gravitational lensing, a fact that must be taken into
account when trying to use this (likely Type Ia) SN to infer
cosmological parameters. Either a more thorough analysis of the
lensing galaxies\footnote{For instance one could try to measure the
  velocity dispersions.} is needed to take the effect into account or
the SN should be used only to set a lower limit on the distance
or maybe even be excluded from the sample if not having large
statistics. In that case some SNe are likely to be de-magnified and the
effect can be assumed to average to $\sim 0$. 
The fact that the intrinsic shape of the host
galaxy looks perfectly normal makes it practically impossible to use this
shape to put constraints on the magnification of SN2003es. If an
extremely unusual shape would have been obtained, it could perhaps
have affected the reliability of the model.   

\section{Summary}
\label{sec:summary}
We have presented the multiple lens-plane method of gravitational
lensing along with its implementation in Q-LET, a \texttt{FORTRAN 77}
code that can be used to quickly estimate the lensing effects on
either a point source or to obtain the intrinsic shape of an
elliptical image. A square grid in the image plane is also possible
for user-defined image shapes.

The code was applied to SN2003es, a SN (most likely Type Ia)
discovered in the HDF-N. Using both NFW and SIS type halos and
estimating the velocity dispersions from the observed luminosities via
the Faber-Jackson and Tully-Fisher relations we found that SN2003es is
likely to have been significantly 
magnified by gravitational lensing and that this problem should be 
addressed in detail before using SN2003es in estimates of the cosmological
parameters. 
 
\ack
The author would like to thank Julio Navarro for permission to use his
code for calculation of the NFW parameters, Phillip Helbig for
permission to use his code for calculations of cosmological
distances, Ariel Goobar, Joakim Edsj\"o and Edvard
M\"ortsell for useful discussions, suggestions and
proofreading. Finally, the author thanks Tomas Dahl\'en. This project was
funded by the Swedish Research Council.

\end{document}